\documentclass[english,twocolumn,PRB, showpacs]{revtex4}
\usepackage{bm}%
\usepackage{amssymb}
\usepackage{amsmath}
\usepackage{graphics}
\usepackage[sort&compress]{natbib}
\usepackage{epsfig}
\usepackage{color}
\usepackage{babel}
\usepackage[ps2pdf,colorlinks=false, pagebackref=false, bookmarks=true,bookmarksopen=true,bookmarksnumbered=true]{hyperref}

\newcommand{\Cu}{\ensuremath{\mathrm{Cu}}}
\newcommand{\Ni}{\ensuremath{\mathrm{Ni}}}
\newcommand{\Nb}{\ensuremath{\mathrm{Nb}}}
\newcommand{\Al}{\ensuremath{\mathrm{Al}}}
\newcommand{\Si}{\ensuremath{\mathrm{Si}}}
\renewcommand{\O}{\ensuremath{\mathrm{O}}}

\newcommand{\Co}{\ensuremath{\mathrm{Co}}}
\newcommand{\Fe}{\ensuremath{\mathrm{Fe}}}
\newcommand{\Pd}{\ensuremath{\mathrm{Pd}}}
\newcommand{\He}{\ensuremath{\mathrm{He}}}

\makeatother

\begin{document}

\title{Josephson tunnel junctions with ferromagnetic $\Fe_{0.75}\Co_{0.25}$ barriers}

\author{D. Sprungmann, K. Westerholt and H. Zabel}

\affiliation{Institut f\"ur Experimentalphysik\:/\:Festk\"orperphysik, Ruhr-Universit\"at Bochum, 44780 Bochum, Germany}

\author{M. Weides and H. Kohlstedt }

\affiliation{Institute of Solid State Research and JARA-Fundamentals of Future Information Technology, Research Centre, J\"ulich, 52425 J\"ulich, Germany}

\begin{abstract}
Josephson tunnel junctions with the strong ferromagnetic alloy $\Fe_{0.75}\Co_{0.25}$ as the barrier material were studied. The
junctions were prepared with high quality down to a thickness range of a few monolayers of Fe-Co. An oscillation length of
$\xi_{F2}\approx 0.79\:{\rm {nm}}$ between $0$ and $\pi$-Josephson phase coupling and a very short decay length
$\xi_{F1}\approx 0.22\:{\rm {nm}}$ for the amplitude of the superconducting pair wave function in the Fe-Co layer were determined.
The rapid damping of the pair wave function inside the Fe-Co layer is caused by the strong ferromagnetic exchange field and
additional magnetic pair breaking scattering. Josephson junctions with
Fe-Co barriers show a significantly increased tendency towards magnetic remanence and flux trapping for larger thicknesses $d_{F}$.
\end{abstract}


\maketitle
\section{Introduction}

The proximity effect in thin film heterostructures combining superconducting
(S) and ferromagnetic (F) layers is a topic of great current interest \cite{buzdin05RMP}. For example, it was shown that the critical temperature $T_c$ depends on the thickness and
magnetic orientation of F-layers in SF bi- or multilayers \cite{Tagirov,ZdravkovReentrantSF,Westerholt}. Another closely related
phenomenon is the occurrence of so-called $\pi$-coupling in SFS-type Josephson junctions (JJs). For certain thickness ranges of the
ferromagnetic layer $d_{F}$ the Josephson coupling energy $E_{J}$ has a minimum for a phase difference $\varphi=\pi$ and not
for $\varphi=0$ as in normal Josephson junctions. The ground state of the Josephson junction oscillates
between $0$- and $\pi$-coupling with the oscillation period given by $\Delta d_{F}=\pi\xi_{F}$. The
magnitude of the magnetic length $\xi_{F}$ depends on the kind of transport regime.  In the clean limit, i.e., when for the electron
mean free path $\ell_m$ in the ferromagnet $\ell_m\gg\xi_{F}$ holds, it is determined by $\xi_{F}=\hbar v_{F}/E_{ex}$, with the Fermi velocity $v_{F}$, the exchange energy $E_{ex}$ and Planck's constant $\hbar$. In
the dirty limit, i.e. if $\ell_m\ll\xi_{F}$, $\xi_{F}$ depends on the electron diffusion constant $D_{F}=\frac{1}{3}v_{F}\ell_m$
via
\[
\xi_{F}=\sqrt{\frac{\hbar D_{F}}{E_{ex}}}\label{xiF}\:.
\]
For SIFS-type junctions in the dirty limit with magnetic scattering the critical current is given by \cite{WeidesHighQualityJJ, VasenkoPRB}:

\begin{eqnarray}
I_{c}R_{\text{n}}(d_{F})\propto\left|\cos{\left(\frac{d_{F}-d_{0}}{\xi_{F2}}\right)}\right|\exp{\left(-\frac{d_{F}}{\xi_{F1}}\right)}\:.
\label{fit}
\end{eqnarray}
where $d_{0}$ denotes the thickness of the non-magnetic (dead magnetic) part of F-layer and I stands for the AlO$_x$ tunnel barrier within the stack,
$R_n$ is the resistance in the normal state.
In general the decay length $\xi_{F1}$ and the oscillation length $\xi_{F2}$ are different. In the limit $E_{ex}\gg k_BT_c$ they are related by the equation \cite{VasenkoPRB}:
\[
\xi_{F1,F2}=\xi_F\sqrt{\frac{1}{\sqrt{1+\left(\frac{\hbar/\tau_{ie}}{E_{ex}}\right)^{2}}\pm\left(\frac{\hbar/\tau_{ie}}{E_{ex}}\right)}}\:,
\label{lengths}
\]
where the positive and negative sign in the denominator refers to $\xi_{F1}$ and $\xi_{F2}$, respectively.

\begin{figure}
\centering
\includegraphics[width=5.6cm]{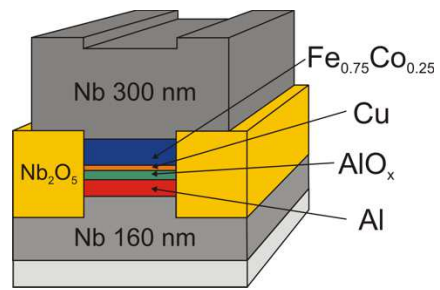}
\caption{(Color online) Schematic design of the SINFS-type junctions, the insulation is prepared by
using an anodic oxidation technique.}
\label{figure01}
\end{figure}

Only if the pair breaking scattering rate $\tau_{ie}^{-1}$ is small, i.e. $\hbar/\tau_{ie}\ll E_{ex}$, $\xi_{F1}\approx\xi_{F2}$ holds.
For strong pair breaking scattering they differ and $\xi_{F1}<\xi_{F2}$.

The experimental confirmation of $\pi$ coupling in SFS-type \cite{Ryazanov01piSFS_PRL,Sellier03TinducedSFS,Blum:2002:IcOscillations} as
well as in SIFS-type JJs \cite{Kontos02Negativecoupling,WeidesHighQualityJJ},
promoted an upsurge of interest in JJs with ferromagnetic barriers. This is motivated by the perspective
of applying $\pi$-coupled Josephson junctions in flux qubits or digital logic circuits \cite{ustinovRapidsingle-fluxquantumlogicusingpi-shifters}.
For example, by combining a tunnel $0$- and a metallic $\pi$-JJ into one superconducting loop,
one can define a quantum mechanical superconducting two level system with
all properties required for the basic unit of quantum computational
devices, so-called qubits \cite{Yamashita:2005:pi-qubit:SFS+SIS}. The physics of fractional vortices can also be studied in junctions with combined $0$ and $\pi$ coupling \cite{WeidesFractVortex}.\\
For the superconducting layer S the elemental superconductor $\Nb$ is nearly
exclusively used in the literature \cite{Sellier03TinducedSFS,Kontos02Negativecoupling,OboznovRyazanov06IcdF,WeidesHighQualityJJ,VavraPRB06,RobinsonPRB07}.
For the F-layer either
diluted ferromagnets like $\Pd\Ni$ \cite{Kontos02Negativecoupling}, $\Ni\Cu$ \cite{OboznovRyazanov06IcdF,WeidesHighQualityJJ} or SiFe \cite{VavraPRB06}
or elemental 3d magnets are applied.
In diluted ferromagnetic alloys the pair breaking of the Cooper pairs
by the exchange field is weak and the oscillation length can be rather large, thus enabling for large thicknesses
$d_{F}$ a still high superconducting critical current density $j_{c}$. This simplifies the observation
of oscillations in $I_{c}(d_F)$, which is taken as the main
experimental evidence for $\pi$-coupling. The first minimum ($0$ to $\pi$) \cite{Kontos02Negativecoupling,WeidesHighQualityJJ}
and also the second minimum ($\pi$ to $0$) \cite{OboznovRyazanov06IcdF} was observed in JJs using Ni-Cu alloys.\\
Junctions with 3d transition metals like $\Fe$, $\Co$ and $\Ni$ as the barrier material were studied in
\cite{RobinsonPRB07,Blum:2002:IcOscillations,Bannykh08}.
By using strong ferromagnets as the barrier material one expects a very short oscillation period, corresponding to a thickness of
a few monolayers only. From the experimental point of view it is challenging to grow homogeneous barriers to obtain junctions with a sufficient comparability, which is crucial to observe the $0$ to $\pi$-transition.
Small lateral inhomogeneities in the ferromagnetic layer would severely deteriorate the junction properties by locally inducing
phase shifts in the oscillating pair density.
The competition between the short oscillation period and the exponentional damping of the critical current density caused by the
strong pair breaking from the exchange field characterizes the shape of the $I_c(d_F)$-dependence. Since
the ratio $\frac{\xi_{F1}}{\xi_{F2}}$ determines the location of the 0-$\pi$-transition, one can estimate the crossover thickness
$d_F^{0-\pi}$ even if the crossover is not observable directly.
The alloy \Fe$_{0.75}$\Co$_{0.25}$, which we chose for the F-layer, has the largest magnetic moment of about $2.5\:\rm{\mu_{B}}$
per atom in the bulk among the 3d transition metal series. Therefore we expect a large exchange energy and correspondingly very short
coherence lengths which should be smaller than those of comparable Fe- or Ni-junctions. The
value of the exchange energy of \Fe$_{0.75}$\Co$_{0.25}$ must be considered as the maximum value possible for SFS/SIFS junctions, based on 3d transition
metals. Until now the composition
\Fe$_{0.75}$\Co$_{0.25}$ has not been studied in the context of $\pi$-junctions.

\section{Preparation and Experimental}

The junctions were fabricated using a combination of dc-magnetron
sputtering and optical lithography, as described in detail in Ref.
\cite{WeidesFabricationJJPhysicaC}. The schematic design of the junctions is depicted
in Fig.\ref{figure01}. The complete thin film stack including the
barrier and a part of the top Nb-electrode was sputtered in one run
in a commercial sputtering chamber (Leybold Univex 450B) with a base
pressure of $4\cdot10^{-7}\:\rm{mbar}$ on a thermally oxidized $\Si$-substrate
at room temperature, see Fig. \ref{figure01}.\\
After sputtering a (Nb(40 nm)Al(2.4 nm))$_{4}$ multilayer, the $5\:\rm{nm}$
thick Al layer was deposited and thermally oxidized for 30 minutes
in pure oxygen at pressures of $1.5\:\rm{ mbar}$ and $0.015\:\rm{ mbar}$, respectively.
By using an oxidation pressure of $0.015\:\rm{ mbar}$ instead of $1.5\:\rm{ mbar}$ the AlO$_x$ layer is definetely thinner and the
critical current density increases by a
factor of 12-15. This is helpfull to improve the signal quality for larger Fe-Co thicknesses, for which we expected a significant supression of the supercurrent.
The purpose of the thin Al interlayers in the Nb/Al multilayer is the optimization of the top Nb/Al interface in order to reduce the barrier rougness to a minimum.
Depending on the oxygen pressure, about $1\textrm{-}2\:\rm{nm}$ of the
$\Al$-layer are transformed into $\Al\O_{x}$. The remaining metallic
$\Al$ film becomes superconducting by the proximity effect below the
transition temperature of $\Nb$. On top of the $\Al\O_{x}$-layer a $2\:\rm{nm}$
thick non-magnetic (N) $\Cu$-film is sputtered, followed by the Fe-Co ferromagnetic film.
In recent work \cite{WeidesFabricationJJPhysicaC} it was shown that the $\Cu$-interlayer is needed to keep the interface
roughness of the F-layer on an acceptable level.
The Fe-Co film is wedge shaped along the substrate length using the natural gradient of the sputtering rate. As a final step
the $\Nb$-counter-electrode with a thickness of $40\:\rm{nm}$ was deposited.\\

After the lift-off process mesas of $10\times50\:\rm{\mu m^{2}}$ and $10\times200\:\rm{\mu m^{2}}$ are defined by optical lithography and ion-beam etching.
Afterwards the current leads and the sides of the stack are isolated by
anodic oxidation of Nb \cite{Kroger81SNAP}. Finally,
after a short Ar plasma etching, the top electrode is completed by
sputtering another 400 nm of Nb.\\
For the present study we prepared two series of junctions with two
different $\Al\O_{\text{x}}$ barrier thicknesses. The first series
with the thinner $\Al\O_{\text{x}}$ barrier covered the thickness range
$d_{F}=0.9\:-\:3.2\:{\rm {nm}}$ for the Fe-Co layer and had a normal state
resistance for the $10\times50\:\rm{\mu m^2}$ JJs of about $R_{\text{n}}=0.16\,$ $\Omega$ and $R_{\text{n}}\approx $ 0.052 $\Omega$
($10\times200\:\rm{\mu m^2}$), respectively.
The second series with a thicker $\Al\O_{\text{x}}$ barrier covered
the thickness range $d_{F}=0.4\:-\:1.1\:{\rm {nm}}$ and had $R_{\text{n}}\approx1.07$
$\Omega$ ($10\times50\:\rm{\mu m^2}$).

The ferromagnetic properties of the thin Fe-Co layers were studied
by a commercial SQUID magnetometer on identical, non-microstructured
layer stacks covering the same Fe-Co thickness range. The I-V characteristics
of the Josephson junctions with and without an applied magnetic field
in the film plane were measured at $T = 4.2\:\rm{K}$ in a shielded $^{4}\He$
cryostat using home made electronics.

\section{Results and discussion}
\begin{figure}
\includegraphics[width=8.6cm]{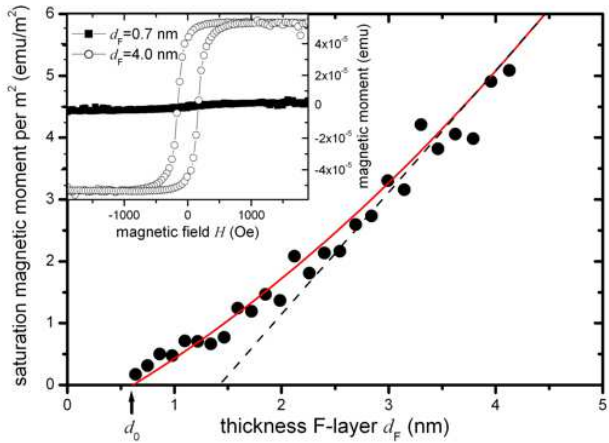}
\caption{Saturation magnetic moment per m$^2$ vs Fe-Co thickness $d_{F}$ measured at $T=15\:{\rm {K}}$.
The dashed line ls a theoretical curve corresponding to a magnetic moment of 2.5 $\mu_B$/atom. The red line leads to a total
thickness  of non magnetic interlayers of $d_0=0.6$ nm. The inset depicts the
hysteresis curves of two samples with different Fe-Co layer thicknesses
$d_{F}=0.7\:{\rm {nm}}$ and $d_{F}=4.0\:{\rm {nm}}$.}
\label{figure04}
\end{figure}

The magnetism of the $\Fe_{0.75}\Co_{0.25}$ alloy layer in the JJs plays
an important role, thus we first characterize the magnetic properties. The magnetic hysteresis loops of the SINFS stacks with
$d_F$ in the same range as for the JJs have been measured at $15\:\rm{K}$. In the inset of
Fig.\ref{figure04} we show two examples
for a thickness $d_{F} = 0.7\:\rm{nm}$ and $d_{F} = 4.0\:\rm{nm}$. For the thicker
film one observes a nearly square shaped hysteresis loop with a coercive
field of $200\:\rm{Oe}$ and a saturation magnetization corresponding to about
70\% of the bulk value. For the thinner film there is only a very
weak ferromagnetic signal, indicating that due to intermixing at the interfaces very thin Fe-Co layers become non-magnetic.
The saturation magnetic moment normalized to the sample area is plottet versus the nominal thickness of the Fe-Co layer in Fig.\ref{figure04}.
Below a nominal thickness of $d_F=0.6$ nm the layers are non ferromagnetic. Above $d_F=0.6$ nm the ferromagnetic moment gradually
increases and approaches the bulk moment of 2.5 $\mu_B$/atom above about $d_F=2.5$ nm. This indicates that at each interface by alloying
with Nb or Cu there is a reduction of the ferromagnetic moment and the first two monolayers at each side are non ferromagnetic.\\

A second experimental ingredient needed for a theoretical description
of the Josephson junctions is the conduction electron mean free path
of the ferromagnetic layer $\ell_m$, which defines whether
the dirty limit or the clean limit theory for the JJs
applies. We measured the electrical conductivity of a single $\Fe_{0.75}\Co_{0.25}$
thin film of $8\:\rm{nm}$ thickness on thermally oxidized $\Si$. The $\Fe_{0.75}\Co_{0.25}$-layer
was capped with 30 nm of sputtered SiO$_2$. The
film had a very small residual resistivity ratio RRR $\approx1.2$
(defined as the ratio of the resistance measured at room temperature
and at $4.2\:\rm{K}$) and a large residual resistivity of $\rho_{m}=27\:\rm{\mu\Omega cm}$.
For Fe-Co in Ref.\cite{kim2006} a smaller value of $\rho_{m}=14.8\:\rm{\mu\Omega cm}$ was determined, indicating
the sensitive influence of different growth conditions.
The electron mean free path can be estimated using the standard free
electron model formula \cite{Pippard60}:

\[
\ell_m=\frac{\pi^{2}k_{B}^{2}}{e^{2}\rho_{m}\gamma_{s}v_{F}}\label{lm}
\]
with the electronic specific heat coefficient $\gamma$ and the Fermi velocity
$v_{F}$. With $\gamma\approx750\:\rm{Jm}^{-3}K^{-2}$ 
and
$v_{F}\approx1.56\cdot10^{6}\:\rm{m/s}$ ($\Co$: $v_{F}=0.28\cdot10^{6}\:\rm{m/s}$, $\Fe$: $v_{F}=1.98\cdot10^{6}\:\rm{m/s}$ \cite{RobinsonPRB07})
we estimated $\ell_m=0.23\, \rm{nm}$. This very small value for the mean
free path indicates strong disorder scattering from the random distribution
of $\Fe$ and $\Co$-atoms in the alloy. The value for $\ell_m$ derived here must be considered as a rough estimate, because
the growth on SiO$_2$ is not directly comparable to the growth within the SINFS-stack.

\begin{figure}[tb]
\includegraphics[width=8.6cm]{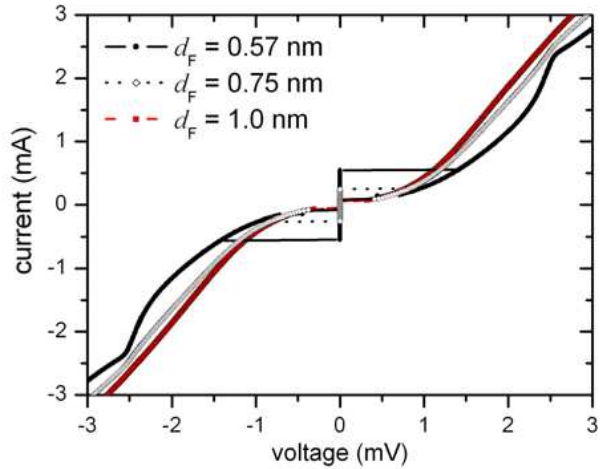}
\caption{(Color online) I-V-curves of three junctions with different Fe-Co thicknesses measured
at $T=4.2\:{\rm {K}}$. The area of the junctions was $10\times50\:\mu\rm{m}^2$.}
\label{figure02}
\end{figure}

In Fig. \ref{figure02} we show I-V-curves of junctions with different
thicknesses of the ferromagnetic layer $d_{F}$. One finds
the typical features of Josephson tunnel junctions with ferromagnetic
barriers \cite{WeidesHighQualityJJ,WeidesFabricationJJPhysicaC,Kontos02Negativecoupling}. For small
thicknesses $d_{F}$ the Josephson phase is strongly underdamped and the I-V-curves exhibit
a pronounced hysteresis. For the lowest thickness in Fig.\ref{figure02}
one can resolve the double superconducting gap $2\Delta$ of $\Nb$ at $2.7 \:\rm{meV}$. With increasing
thickness $d_{F}$ this feature and the hysteresis gradually
vanish.

\begin{figure}[tb]
\includegraphics[width=8.6cm]{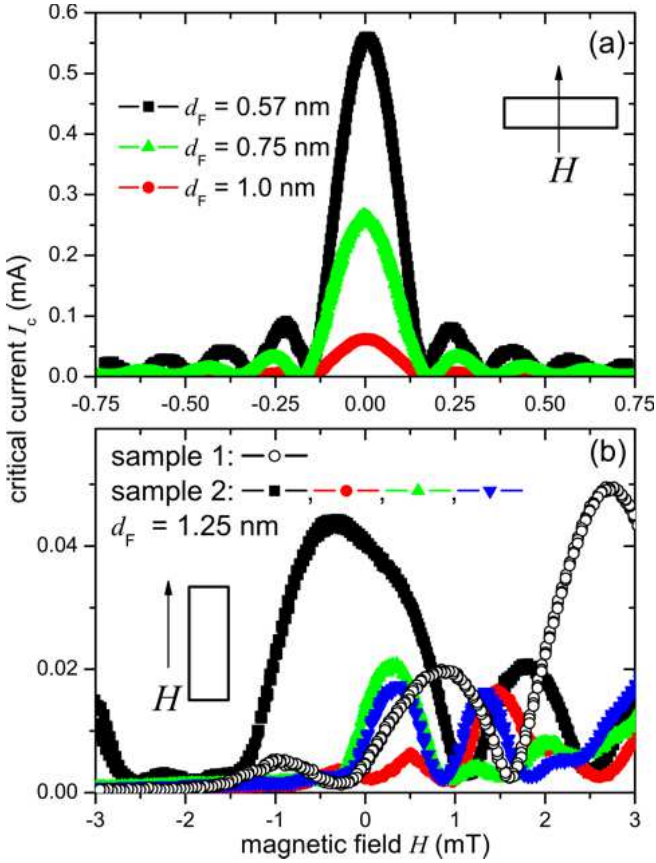}
\caption{(Color online) {\bf{a) $H\perp$ long axis:}} In the thickness range $d_{F}< 1.1\:\rm{nm}$ the
Fraunhofer patterns exhibit mirror symmetry.
{\bf{b) $H\parallel$ long axis:}} For other junctions with thicknesses $d_{F}\geq 1.1\:\rm{nm}$ flux trapping effects become inevitable.
The solid symbols belong to one junction measured for various cooling cycles ($\Delta T \approx 200$ K). The junctions were cooled
without applying an external magnetic field.
The open circles belong to a sample with a similar thickness $d_F$ showing the shift of the Fraunhofer pattern exemplarily.
In both cases a) and b) the sample size was $10\times50\:\rm{\mu m^2}$. }
\label{figure03}
\end{figure}

\begin{figure}[tb]
\includegraphics[width=8.6cm]{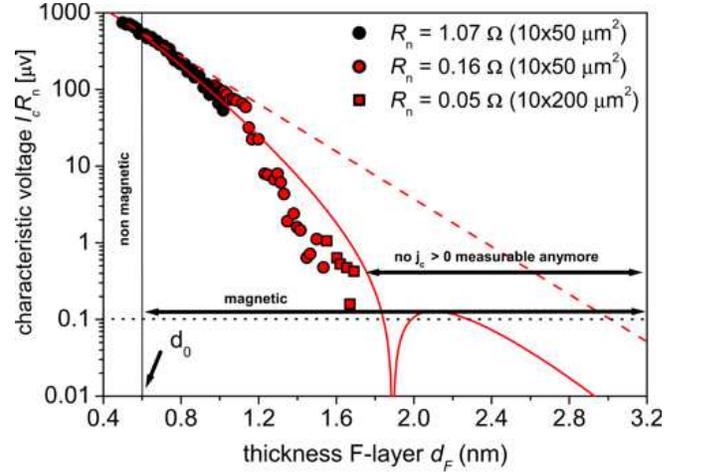}
\caption{(Color online) $I_{c}R_{\text{n}}$ product vs Fe$_{0.75}$Co$_{0.25}$ thickness $d_{F}$.
The solid line is a theoretical curve corresponding to Eq. \ref{fit},
the dashed curve corresponds to Eq.\ref{efit} for thicknesses $d_{F}\leq  0.9 \:\rm{nm}$. The dotted line marks the resolution
limit of our instrumental setup. The parameters for the theoretical curve are: $E_{ex}= 453\:\rm{meV}$, $\xi_{F1} = 0.22 \:\rm{nm}$,
$\xi_{F2} = 0.79\:\rm{nm}$, $\xi_{F}= 0.42\:\rm{nm}$, $\ell_m = 0.23\:\rm{nm}$, $\hbar/\tau_{ie} = 1.66\:E_{ex}$.
In order to compare both series with different AlO$_x$ thicknesses the characteristic voltage $I_cR_{\text{n}}$ is plotted
against $d_F$. The squares and the circles are data points from samples which were produced within the same run. For thicknesses
$d_{F}\geq 1.7$ nm no critical currents could be observed anymore.}
\label{figure05}
\end{figure}

The Fraunhofer pattern $I_{c}(H)$ of the same junctions are shown
in Fig.\ref{figure03}a. For $d_{F}<1.1\:{\rm {nm}}$ we observe
highly symmetric, periodic patterns with vanishing critical current
at the minima, which is a clear indication of the homogeneity of
both the $\Al\O_{\text{x}}$ and Fe-Co barriers in our samples. For thicknesses $d_{F}>1.1\:{\rm {nm}}$
(see Fig.\ref{figure03}b) the global maximum of the Fraunhofer patterns
was found to be increasingly shifted towards external fields $H\neq0$,
indicating the existence of an intrinsic magnetic stray field. Similar observations in $\Ni$-based JJs have been published
recently in Ref.\cite{WeidesAnisotropySIFS}. In the case of Ni the flux trapping effects appear for thicknesses $d_F\geq 3.7\:\rm{nm}$ which
is significantly larger than in our samples. In the case of Fe$_{0.75}$Co$_{0.25}$ the magnetic stray field appears within the 0-state, whereas in the case of
Ni the samples are already in the $\pi$-state.
The $I_c(H)$-curve of one sample (open circles) in Fig.\ref{figure03}b appears to
be shifted by $\sim 2.5\:\rm{ mT}$ along the field axis, corresponding to $2\textrm{-} 3\:\Phi_{0}$.
In this case the external field was parallel to the long axis of the junction. If $H$ was orientated perpendicular to the long axis
only extremely small critical currents could be measured even for maximum fields around 3 mT. This indicates the presence
of a magnetic shape anisotropy within the F-layer, which shifts the Fraunhofer pattern opposite to the magnetization direction \cite{WeidesAnisotropySIFS}.
Additionally, we usually found indication of some trapped magnetic flux, probably emanating from the ferromagnetic film and frozen
in below $T_{c}$. This flux leads to irregular, not reproducible $I_{c}(H)$ patterns and the global maximum critical current
$I_{c}$ cannot be defined precisely. The maximum experimental
critical current which is seen in the experiment is always an underestimation for the true maximum critical current. Additionally,
this random error causes an increasing scattering of the data
points for $d_{F}\geq 1.1 \:\rm{nm}$ in $I_{c}(d_{F})$.\\
In Fig.\ref{figure05} the global maximum critical current $I_{c}$ determined by $I_c(H)$ versus the nominal
thickness of the ferromagnetic layers $d_{F}$ is depicted. In fact the product of the critical current
and the normal state junction resistance $I_{c}R_{\text{n}}$ is plotted for a better comparison of the two series of samples
with different thickness of the $\Al\O_{\text{x}}$-barrier.

$I_{c}$ is strongly damped for increasing $d_{\text{F}}$ and reaches our experimental resolution limit
of $I_{c}R_{\text{n}}\approx 0.1 \:\rm {\mu V}$ beyond $d_{F} > 1.7 \:\rm{nm}$.\\
For a quantitative analysis of $I_{c}R_{\text{n}}(d_{\text{F}}$)
we divided the data into two sections, as done in case of $\Ni$-based SIFS junctions in Ref. \onlinecite{Bannykh08}.
For $d_F\leq 0.6\:\rm{nm}$ the JJs are approximately treated as a SINS-type junction, with
$E_{ex}=0$ meV. Then Eq.\ref{fit} converts into:

\begin{eqnarray}
I_{c}R_{\text{n}}(d_{F})\propto \exp{\left(-\frac{d_{F}}{\xi_{N}}\right)}\label{efit}\end{eqnarray}

with the damping length

\begin{eqnarray}
\xi_{N}=\sqrt{\frac{\hbar D_{N}}{2(\pi k_{B}T+\frac{\hbar}{\tau_{ie}})}}\label{xin}
\end{eqnarray}

and the diffusion constant in the non-ferromagnetic layer $D_{N}$ (see Ref.\cite{VasenkoPRB}).
From the tangent at very thin Fe-Co layers ($d_{F}\leq 0.6 \:\rm{nm}$) (see
Fig.\ref{figure05}, dashed line) we derive a very short damping length
$\xi_{N}\approx 0.28\:\rm{nm}$ and $I_{\text{c}}R_{\text{n}}(d_{F}=0.6\:\text{nm})\approx 1 \rm{m V}$,
corresponding to a magnetic scattering energy of $\hbar/\tau_{ie}\approx
500\:\rm{meV}$ and a critical current density of the SINS stack of $j_{c}= 200\:\rm{A/cm^{2}}$. This
indicates a decrease of $I_cR_n$ by a factor of 2 caused by the F-layer interface scattering compared to SIS or SINFS JJs. In the case
of SIFS JJs with the magnetically weaker $\Ni$ as interlayer, a decrease by factor of $\sim5$ was determined \cite{Bannykh08}.

In the thickness range $d_{F}>d_0=0.6$ nm the ferromagnetism sets in
(see Fig.\ref{figure03}) and the theory for SIFS junctions applies. From the envelope of the function $I_c(d_F)$ following
Eq.\ref{fit}, which was applied to our
data points in the thickness range between $d_{F}= 0.6\:\textrm{-}\:1.7\:\rm{nm}$, we can estimate the damping
length $\xi_{F1}\approx 0.22 \:\rm{nm}$ (see Fig.\ref{figure05}). Obviously, the damping in the magnetic part of the film is stronger than
in the non-magnetic part ($\xi_{N} = 0.28 \:\rm{nm}$). This indicates the effect of the strong exchange interaction on the
pair density.

For a quantitative comparison with theory the model for SIFS junctions can be adapted to our system by considering the IN-layers between
Nb and Fe-Co as one interface with a very low transparency. In this model the oscillation length $\xi_{F2}=
0.79 \:\rm{nm}$ is definitely larger then the decay length $\xi_{F1} = 0.22\:\rm{nm}$, indicating a considerable influence of pair breaking scattering on the tunneling.
Using the experimental value for $\xi_{F1}$, $\xi_{F2}$
and $\ell_m = 0.23 \:\rm{nm}$ we calculate $E_{ex} = 453 \:\rm{meV}$ and $\hbar/\tau_{ie}= 750 \:\rm{meV}$.
The ratio $\frac{\hbar/\tau_{ie}}{E_{ex}}=1.66$ is slightly larger but has the same magnitude as the corresponding value of $\Ni\Cu$ where
$\hbar/\tau_{ie}\approx 1.33\cdot E_{ex}^{NiCu}\approx 100 \:\rm{meV}$ has been determined \cite{OboznovRyazanov06IcdF}. Obviously the
scattering energy of the $\Ni\Cu$-alloy is 7.5 times smaller than the one for Fe-Co.
The value obtained for the exchange energy $E_{ex} = 453 \:\rm{meV}$ is found to be significantly larger than
for the elemental magnets $\Ni$ ($80 - 200 \:\rm{meV}$), $\Co$ ($309\:\rm{meV}$) or $\Fe$ ($256 \:\rm{meV}$)
\cite{RobinsonPRB07}. This seems reasonable considering the strong magnetic properties of Fe$_{0.75}$Co$_{0.25}$.

It should be mentioned, that although the 0 to $\pi$-transition is
located below the instrumental resolution limit of $0.1\: \rm{\mu V}$, the thickness dependence in Fig.\ref{figure05} suggests
a crossover $0-\pi$ at $d_F\approx 1.9$ nm. The experimental curve cannot be fitted reasonably by a pure exponentional decay, a fit with Eq.\ref{fit} is much better.
The calculated oscillation period $\Delta d_F\:=\:2.48\:\text{nm}$ fits perfectly into the series of values obtained for
other systems like $\Delta d_F\:=\:3.7\:\text{nm}$ \cite{Bannykh08} for pure Ni and $\Delta d_F\:=\:11.0\:\text{nm}$
\cite{OboznovRyazanov06IcdF} for the Ni$_{0.53}$Cu$_{0.47}$ alloy.

\section{Summary and conclusions}

We have shown that high quality JJs with the strong
ferromagnetic alloy Fe-Co as the barrier layer can be grown. The high
quality of the barriers has been demonstrated by the observation of
excellent Fraunhofer patterns in case of non-magnetic or magnetically weak,
i.e. very thin F-layers. The JJs are
characterized by a strong damping of the superconducting pair amplitude
in the Fe-Co layers and a small oscillation period for the transition
between 0-coupling and $\pi-$coupling. The strong damping of the
pair amplitude even for the thickness range below $d_{F}=0.6\:{\rm {nm}}$,
where the Fe-Co layers appear to be non-ferromagnetic in the hysteresis loops,
is probably caused by pair breaking scattering on magnetic fluctuations,
which exist in itinerant ferromagnets close to a ferromagnetic phase
boundary. The ferromagnetic range, too, is characterized by strong
inelastic pair breaking scattering of the Cooper pairs, indicating
a high density of states for low energy magnetic excitations in the
Fe-Co ferromagnetic layer with a thickness of only a few monolayers.
Additionally we observe non-reproducible flux trapping
and intrinsic magnetization effects.\\
Applying the dirty limit model to our data gives a reasonable set of parameters and a consistent picture of our system.
Although the 0 to $\pi$-transition of the JJ was not directly observable due
to the strong damping of the critical current density $j_c$ we could estimate its position at $d_{F}\approx 1.9\:\rm{nm}$.

\subsection*{Acknowledgement}

The authors thank R. Waser, G. Pickartz and R. Borowski for support. D. S., K. W. and H. Z. acknowledge financial support
by SFB-491, and M. W. by project WE 4359/1-1.

\end{document}